\documentclass[12pt]{article}
\usepackage{graphicx}

\newcommand\pubnumber{SNSN-XXX-YY}
\newcommand\pubdate{\today}

\def\Title#1{\begin{center} {\Large #1 } \end{center}}
\def\Author#1{\begin{center}{ \sc #1} \end{center}}
\def\Address#1{\begin{center}{ \it #1} \end{center}}

\newcommand\pubblock{\rightline{\begin{tabular}{l} \pubnumber\\
         \pubdate  \end{tabular}}}
\newenvironment{Abstract}{\begin{quotation}  }{\end{quotation}}
\newenvironment{Presented}{\begin{quotation} \begin{center} 
             PRESENTED AT\end{center}\bigskip 
      \begin{center}\begin{large}}{\end{large}\end{center} \end{quotation}}

\begin{document}
\begin{titlepage}
\pubblock

\vfill
\Title{Search for Neutrinoless Double-Beta Decay}
\vfill
\Author{Werner Tornow \footnote{Talk given on behalf of the KamLAND-Zen Collaboration.}}
\Address{Department of Physics, Duke University, Durham NC, 27708, USA \\
        and \\
    Triangle Universities Nuclear Laboratory, Durham, NC, 27708, USA \\
        and \\
    Kavli Institute for the Physics and Mathematics of the Universe, University of
        Tokyo, Kashiwa, 277-8583, Japan}

\vfill
\begin{Abstract}
After the pioneering work of the Heidelberg-Moscow (HDM) and International Germanium
Experiment (IGEX) groups, the second round of neutrinoless double-$\beta$ decay searches
currently underway has or will improve the life-time limits of double-$\beta$ decay
candidates by a factor of two to three, reaching in the near future the $T_{1/2} = 3 \times
10^{25}$ yr level. This talk will focus on the large-scale experiments GERDA, EXO-200, and
KamLAND-Zen, which have reported already lower half-life time limits in excess of
$10^{25}$ yr. Special emphasis is given to KamLAND-Zen, which is expected to approach the
inverted hierarchy regime before future 1-ton experiments probe completely this life-time
or effective neutrino-mass regime, which starts at $\approx 2 \times 10^{26}$ yr or
$\approx 50$ meV.  
\end{Abstract}
\vfill
\begin{Presented}
XXXIV Physics in Collision Symposium \\
Bloomington, Indiana,  September 16--20, 2014
\end{Presented}
\vfill
\end{titlepage}
\def\thefootnote{\fnsymbol{footnote}}
\setcounter{footnote}{0}

\section{Introduction}
Neutrinoless double-$\beta$ decay ($0\nu\beta\beta$) provides the only known experimental
approach for testing the Majorana nature of neutrinos, i.e., finding out whether neutrinos
are their own anti-particles. The observation of $0\nu\beta\beta$ implies that neutrinos
are Majorana particles and that lepton number conservation is violated. It also gives
information on the absolute mass scale of neutrinos, which cannot be obtained from
neutrino oscillation experiments. A Majorana mass term would provide an explanation for
the lightness (compared to the other Standard Model leptons) of neutrinos via the seesaw
mechanism. Finally, Majorana neutrinos may also give the answer to the question why there
is an excess of matter over antimatter in the observable universe. 

The decay rate for $0\nu\beta\beta$ is given by 
\begin{equation}
(T_{1/2}^{0\nu})^{-1} = G_{0\nu} |M_{0\nu}|^2 (\langle m_{\beta\beta}\rangle /m_e)^2,  
\end{equation}
where $G_{0\nu}$ is a phase-space factor (including coupling constants), $M_{0\nu}$ is the
nuclear matrix element (NME) involved, $m_e$ is the electron mass, and $\langle
m_{\beta\beta} \rangle $ is the effective Majorana neutrino mass given by
\begin{equation}    
\langle m_{\beta\beta} \rangle =\bigg| \sum\limits_{i=1}^{i=3}U_{ei}^2 m_i\bigg| .
\end{equation}
Here, $U_{ei}$ is the admixture of neutrino mass eigenstate $i$ in the electron neutrino.

\section{GERDA}

\begin{figure}[htb]
  \centering
  \includegraphics{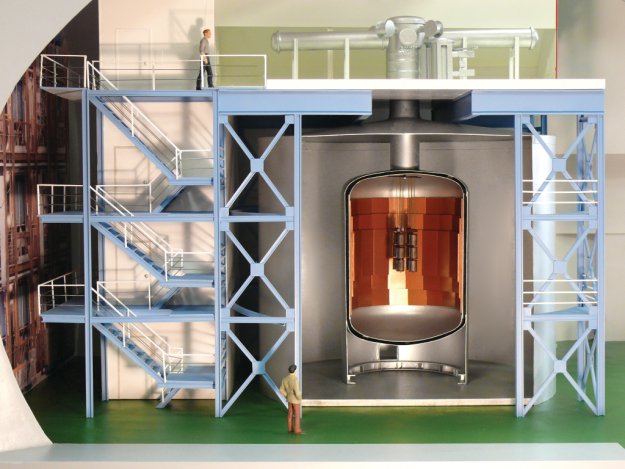}
  \caption{Schematic of the GERDA $0\nu\beta\beta$ decay setup showing the liquid argon filled
  cryostat with an array of HPGe detectors at the center. }
  \label{fig1}
\end{figure}

The Germanium Detector Array (GERDA) collaboration published the result of their Phase I
experiment in 2013 \cite{AGO13}. The experiment is located at the Gran Sasso Laboratory LNGS in
Italy, and it uses germanium both as source and detector for the two electrons emitted in
double-$\beta$ decay. High-purity germanium (HPGe) detectors enriched to ~86\% in $^{76}$Ge
are mounted in low-mass copper supports which
are immersed in a 64 m$^3$ cryostat filled with liquid argon. The liquid argon serves two
purposes: it is the cooling medium for the HPGe detectors and it shields them from
external background radiation. A water Cherenkov detector acts as muon veto. A schematic
view of the GERDA setup is shown in Fig.~\ref{fig1}. 

Reprocessed HPGe detectors from the HDM \cite{KLA01} and IGEX \cite{AAL02, AAL04}
experiments and Canberra type broad
energy germanium (BEGe) detectors were used. The total exposure was 21.6 kg yr of enriched
Ge detector mass. 

\begin{figure}[htb!]
  \centering
  \includegraphics[width=4.0in]{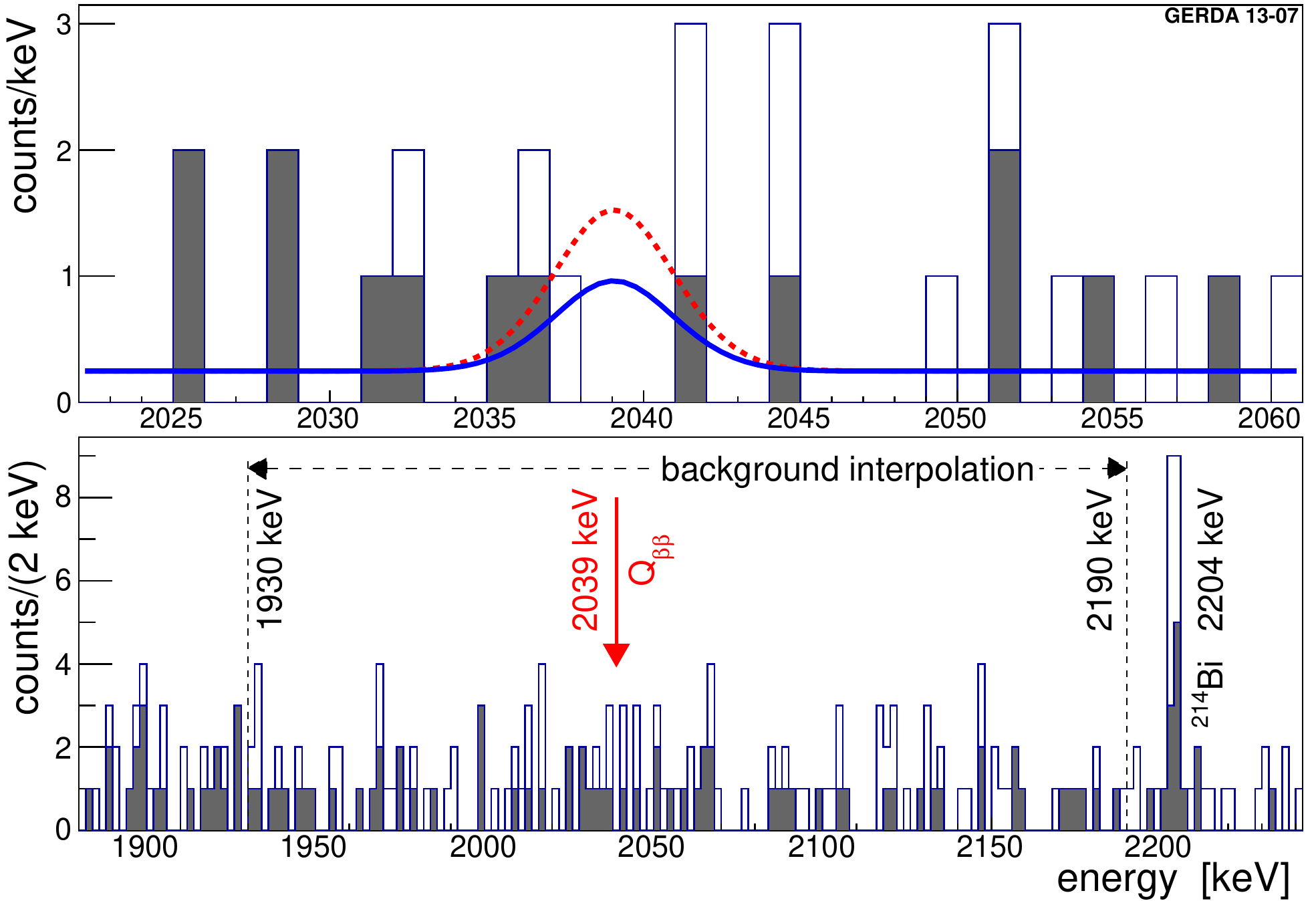}
  \caption{GERDA energy spectra \cite{AGO13}. Lower panel: Energy spectrum centered at the
  Q-value of $0\nu\beta\beta$ of  $^{76}$Ge. Open
  histogram without PSD and solid histogram with PSD cut applied. Upper panel: Expanded
  view of the region of interest with prediction based on the claimed discovery \cite{KLA04,
  KLA06} of $0\nu\beta\beta$ of $^{76}$Ge (dotted curve) and upper limit obtained by the GERDA
  collaboration (solid curve).}
  \label{fig2}
\end{figure}

The combined energy spectrum from all detectors is shown in Fig.~\ref{fig2}. The open histogram is
without pulse-shape discrimination (PSD) constraints applied, while the filled histogram
includes PSD. The upper panel displays the region of interest centered at 2039 keV, the
Q-value for $0\nu\beta\beta$ of $^{76}$Ge, while the bottom panel gives the energy region
used for background determination. Clearly, there are no events in the energy region of
interest. The dotted curve in the upper panel is based on the expectation using the
central value of $T_{1/2} = 1.19 \times 10^{25}$ yr claimed by a subset of the HDM
collaboration \cite{KLA04, KLA06}. The solid curve is the upper limit obtained by the GERDA
collaboration \cite{AGO13}. Their limit is 
$T_{1/2}^{0v} > 2.1 \times 10^{25}$ yr (90\% C.L.)
The quoted background in the energy region of interest is $10^{-2}$ counts
keV$^{-1}$kg$^{-1}$yr$^{-1}$ after the PSD cut is applied. This value is about a factor of 10
larger than future 1 ton experiments are aiming for. 
The GERDA collaboration combined their result with those of the HDM \cite{KLA01} and IGEX
\cite{AAL02, AAL04} experiments and obtained the new limit of
$T_{1/2}^{0\nu} > 3.0 \times 10^{25}$ yr (90\% C.L.).
This value corresponds to an upper limit on the effective neutrino mass range of 200 to 400
meV, using the NME calculations of \cite{SIM13, MUS13, ROD10, MEN09,
BAR13, SUH10, MER13} and the phase-space factor of \cite{KOT12}.

\section{EXO-200}

\begin{figure}[htb]
  \centering
  \includegraphics[width=4in]{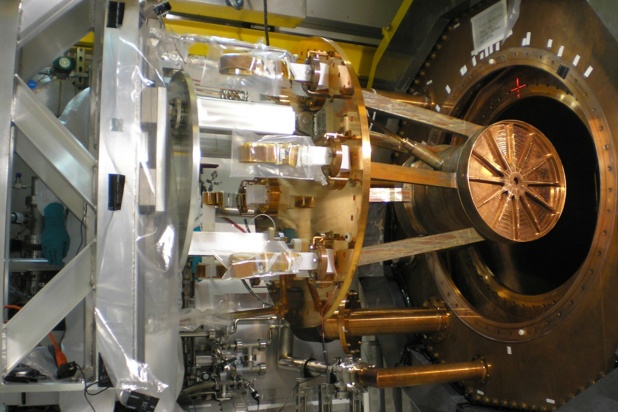}
  \caption{Photograph of the EXO-200 cryostat housing the liquid xenon TPC. }
  \label{fig3}
\end{figure}

The EXO (Enriched Xenon Observatory)-200 experiment \cite{AUG12} uses xenon both as source and
detector. It is located at a depth of approximately 1580 meter water-equivalent at the
Waste Isolation Pilot Plant (WIPP) near Carlsbad, New Mexico. The detector consists of a
cylindrical homogeneous time-projection chamber (TPC) filled with liquefied xenon enriched
to 80.6\% in $^{136}$Xe ($0\nu\beta\beta$ decay Q-value of 2457.8 keV). Energy deposited by
charged particles in the TPC produces both ionization and scintillation signals. The TPC
provides three-dimensional topological and temporal information which allows for the
reconstruction of individual energy depositions. The cylindrical TPC is divided into two
symmetric volumes which are separated by a cathode grid. Wire grids installed near both
ends of the TPC serve as charge induction and charge collection grids for deriving the
spectroscopic information. Finally, at each end of the TPC approximately 250 Large Area
Avalanche Photodiodes collect the scintillation light. The TPC is installed at the center
of a low-background cryostat, as shown in Fig.~\ref{fig3}.

\begin{figure}[htb]
  \centering
  \includegraphics[width=4.5in]{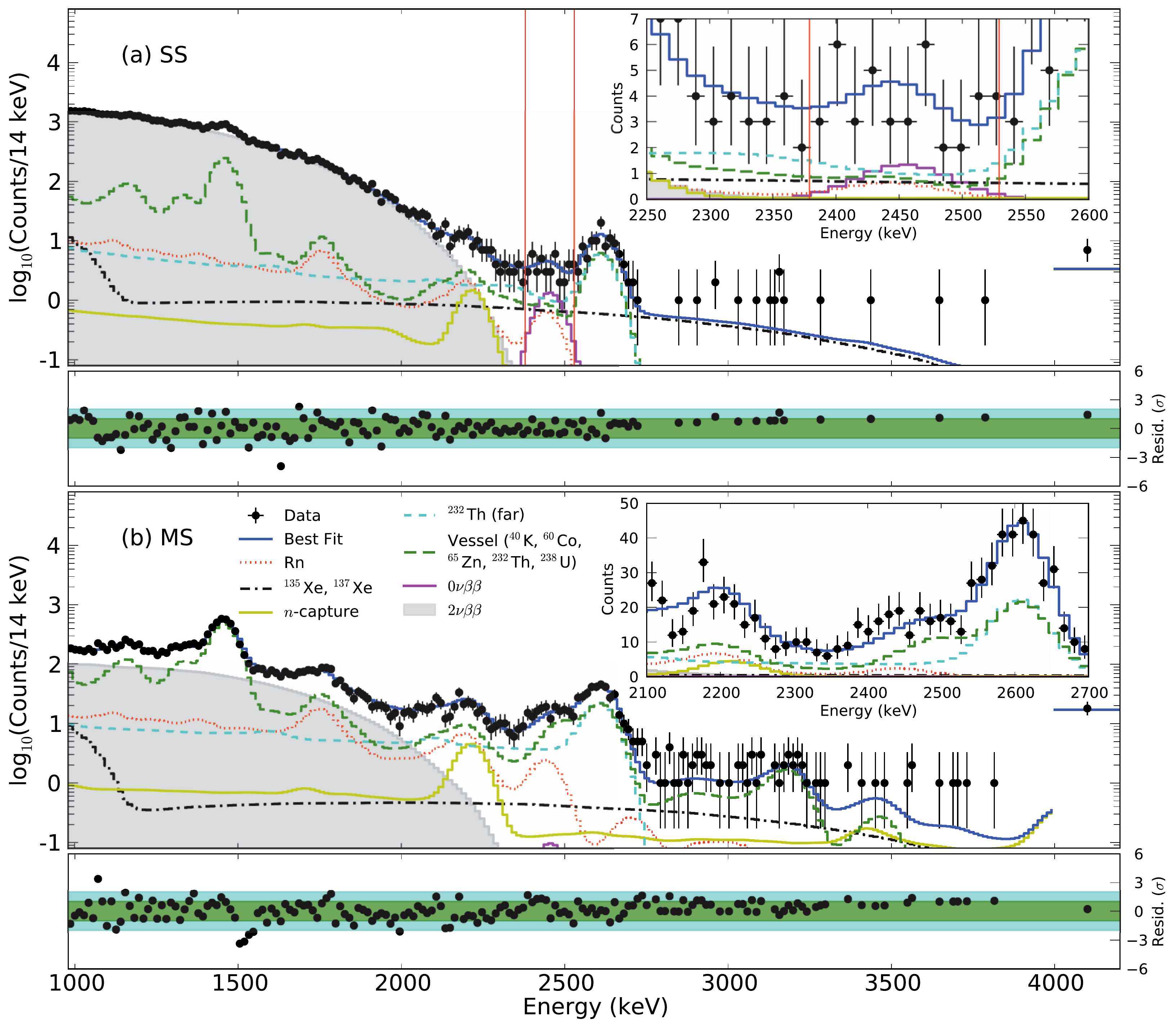}
  \caption{EXO-200 energy spectra \cite{ALB14}. Fit results projected onto energy axis.
  Main panel a (top) shows single-site
  (SS) events and main panel b (bottom) shows multi-site (MS) events versus energy. The
  data are shown as dots with error bars. The insets are zoomed-in displays of the region
  of interest for $0\nu\beta\beta$ indicated in the main panel a by the vertical lines. The lower
  panels in a and b show residuals between data and best fit.  }
  \label{fig4}
\end{figure}

The number of charge deposits allows for a classification of the events into two groups.
The $0\nu\beta\beta$ decay events of interest are Single-Site (SS) events, while
background events are mostly Multi-Site (MS) events. The maximum–likelihood fit published
by the EXO-200 collaboration is shown in Fig.~\ref{fig4} \cite{ALB14}. It is based on a total xenon
exposure of 123.7 kg yr. The top panel a is for SS events, while the bottom panel b is for
MS events. The shaded area in both panels is a fit to $2\nu\beta\beta$ events. The region
of interest for $0\nu\beta\beta$ is shown in panel a by the vertical lines centered around
2458 keV. The inset in panel a and panel b gives a zoomed-in view of the energy region of
interest. The best-fit model provides an estimate for the background in the
$0\nu\beta\beta$ $\pm 2\sigma$ region of 31.1 $\pm$ 1.8 (stat.) $\pm$ 3.3 (syst.) counts,
which corresponds to $(1.7 \pm 0.2) \times 10^{-3}$ keV$^{-1}$kg$^{-1}$yr$^{-1}$, about a
factor of six smaller than obtained by the GERDA collaboration for $^{76}$Ge. The largest
backgrounds are from $^{232}$Th (16.0 counts), $^{238}$U (8.1 counts), and $^{137}$Xe (7.0
counts). The reported lower limit for the $0\nu\beta\beta$ decay half-life time is
\cite{ALB14}
$T_{1/2}^{0\nu}= 1.1 \times 10^{25}$ yr   (90\% C.L.). 
Using the nuclear matrix elements of \cite{ROD10, MEN09, BAR13, SIM13} and the
phase-space factor of \cite{KOT12} gives an upper limit for the Majorana neutrino mass of 190 -
450 meV.  Finally, the EXO-200 collaboration reports a sensitivity of $1.9 \times 10^{25}$
yr, an improvement by a factor of 2.7 over their earlier result published in \cite{AGO13}. 

\section{KamLAND-Zen}

\begin{figure}[htb]
  \centering
  \includegraphics[width=4.0in]{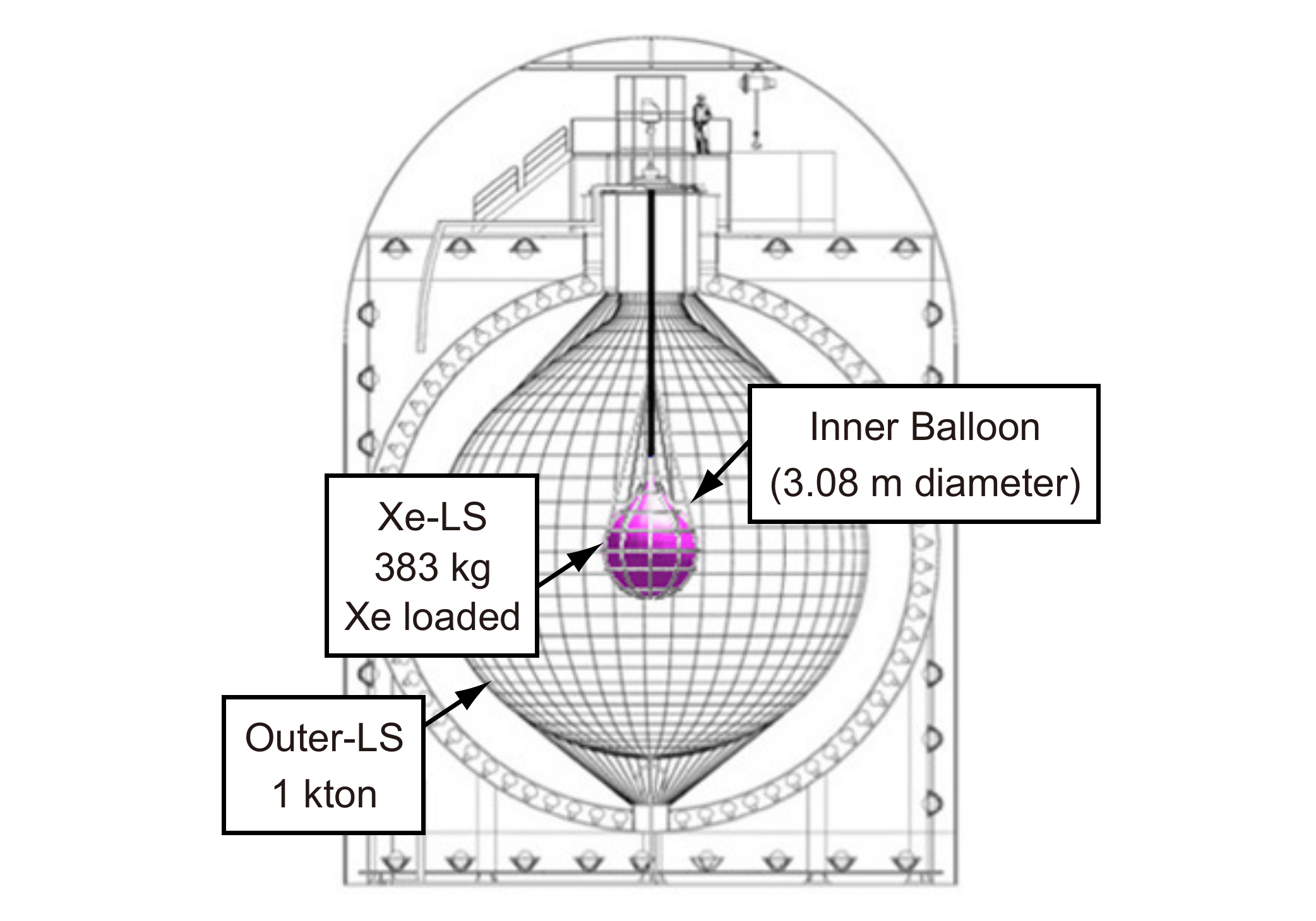}
  \caption{Schematic view of KamLAND-Zen detector \cite{ASA14} consisting of xenon
  (enriched to 91\% in $^{136}$Xe) loaded liquid
  scintillator inner balloon surrounded by outer balloon filled with regular liquid
  scintillator and viewed by PMTs mounted to the inner surface of a 18 m diameter
  stainless-steel sphere which in turn is surrounded by a water-Cherenkov veto detector. }
  \label{fig5}
\end{figure}

The KamLAND-Zen (Kamioka Liquid Scntillator Anti-Neutrino Detector-Zero neutrino)
double-$\beta$ decay experiment \cite{GAN13} consists of 13 tons of Xe-loaded (320 kg of
$^{136}$Xe) liquid scintillator (Xe-LS) contained in a transparent nylon inner balloon
(IB, also denoted as mini balloon) of 3.08 m diameter, suspended at the center of the
KamLAND detector by straps. As can be seen from Fig.~\ref{fig5}, the IB is surrounded by 1 kton of
liquid scintillator (LS) contained in an outer balloon of 13 m diameter, which in turn is
surrounded by buffer oil. Photomultiplier tubes (PMTs) providing 34\% photocathode
coverage are mounted on the inner surface of the stainless-steel containment tank to
detect light produced by charged particles in the Xe-LS and LS. The spherical tank is
surrounded by a 3.2 kton water-Cherenkov detector with PMTs attached to its top, bottom,
and cylindrical surface for cosmic-ray muon identification.

\begin{figure}[htb!]
  \centering
  \includegraphics[width=2.5in, angle=270]{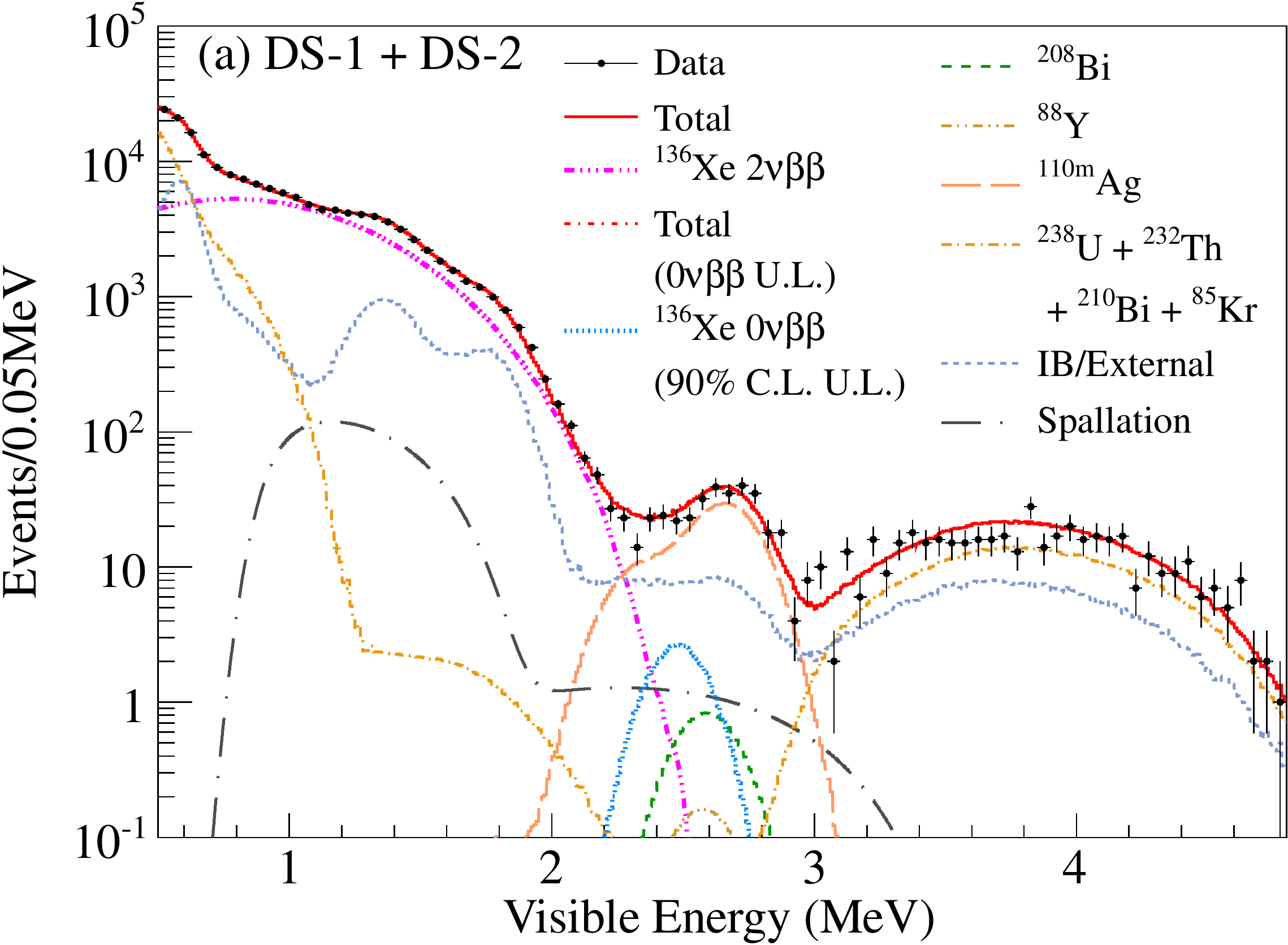}
  \includegraphics[width=2.5in, angle=270]{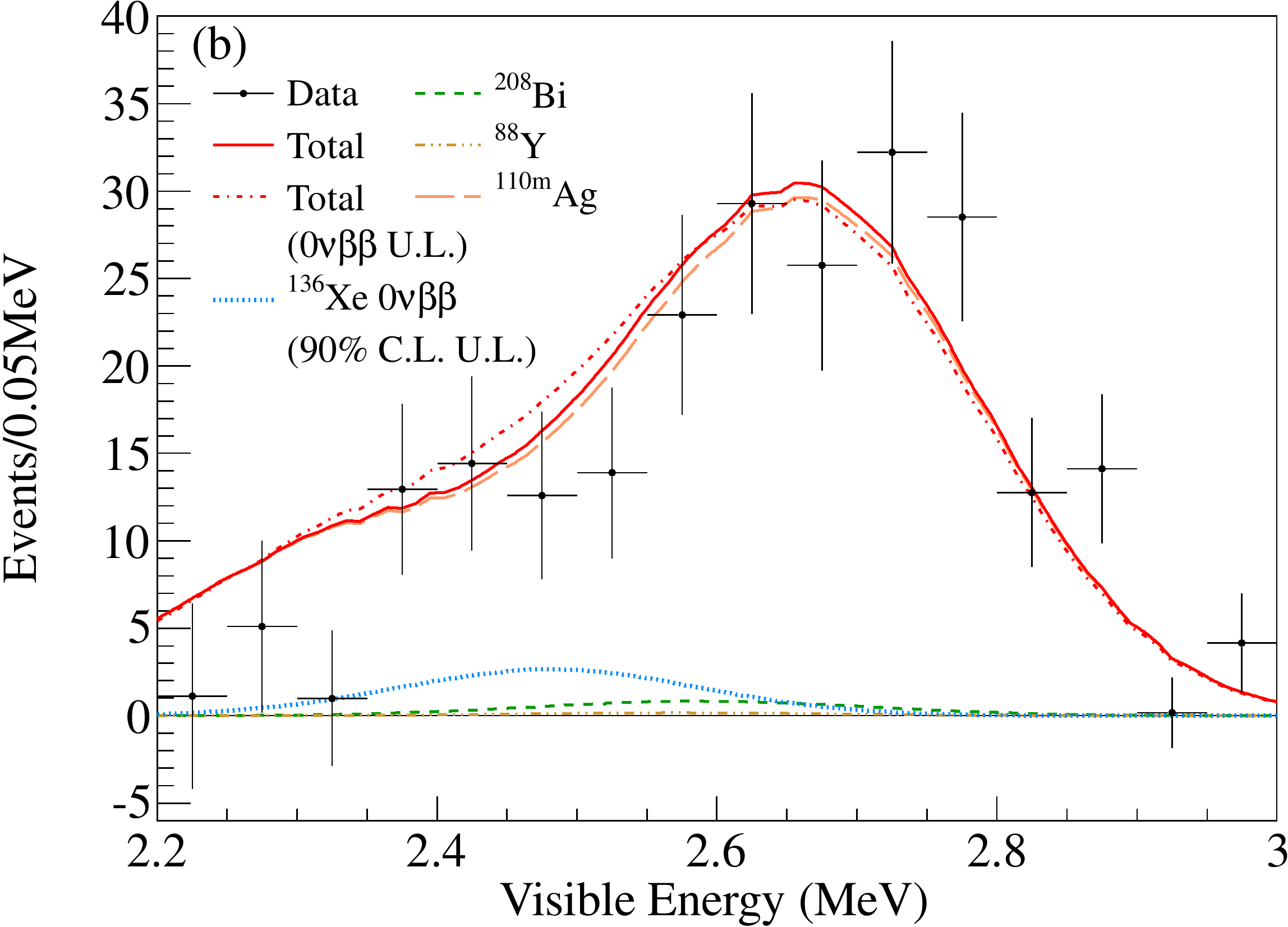}
  \caption{KamLAND-Zen Phase I energy spectra \cite{ASA14}. Top panel (a): Energy spectrum
  together with best-fit backgrounds and $2\nu\beta\beta$
  decay events. Bottom panel (b): Zoomed-in version of  (a) with focus on the $2.2 < E
  <3.0$
  MeV region  after subtracting known background contributions.   }
  \label{fig6}
\end{figure}

Phase I of KamLAND-Zen data collection commenced in 2011. After an exposure of 89.5 kg yr
of $^{136}$Xe, the collaboration reported a lower limit for the $0\nu\beta\beta$ half-life time
of $T_{1/2}^{0\nu} > 1.9 \times 10^{25}$ yr (90\% C.L.) \cite{GAN13}.The sensitivity in the
Phase I search was limited by background resulting from the decay of $^{110m}$Ag. This
contamination produced a peak in the visible energy spectrum (at $\approx$2.6 MeV) which
overlapped with the region of interest for $0\nu\beta\beta$ of $^{136}$Xe, as can be seen
from Fig.~\ref{fig6} \cite{GAN13}. 

In June of 2012 a purification campaign was initiated. First, xenon was extracted from the
IB and it was subsequently found that the $^{110m}$Ag remained in the xenon depleted LS.
Unfortunately, a pump used for Xe-LS extraction developed a leak, resulting in radioactive
impurities in the circulating LS, which accumulated at the bottom part of the mini
balloon. The extracted xenon and newly acquired xenon were purified via distillation and
adsorption, while the LS was purified through water extraction and distillation. However,
it turned out that the entire purification effort resulted in only a factor of three to
four reduction of $^{110m}$Ag. As a consequence, the collaboration decided to start a
time-consuming LS   purification effort involving three volume exchanges in circulation
mode. The extracted and purified xenon was dissolved into the newly purified LS in
November of 2013. In December 2013 data-acquisition of Phase II began, indicating a factor
of approximately 10 in reduction of $^{110m}$Ag.

%\begin{figure}[htb]
%  \centering
%  \includegraphics[width=4.0in]{fig7.pdf}
%  \caption{KamLAND-Zen Phase-II preliminary energy spectra of selected double-$\beta$ candidates within a radius
%  cut of $R<1.0$ m \cite{ASA14}. The residuals from the best fit are shown in the upper panel.}
%  \label{fig7}
%\end{figure}

\begin{figure}[htb]
  \centering
  \includegraphics[width=4in]{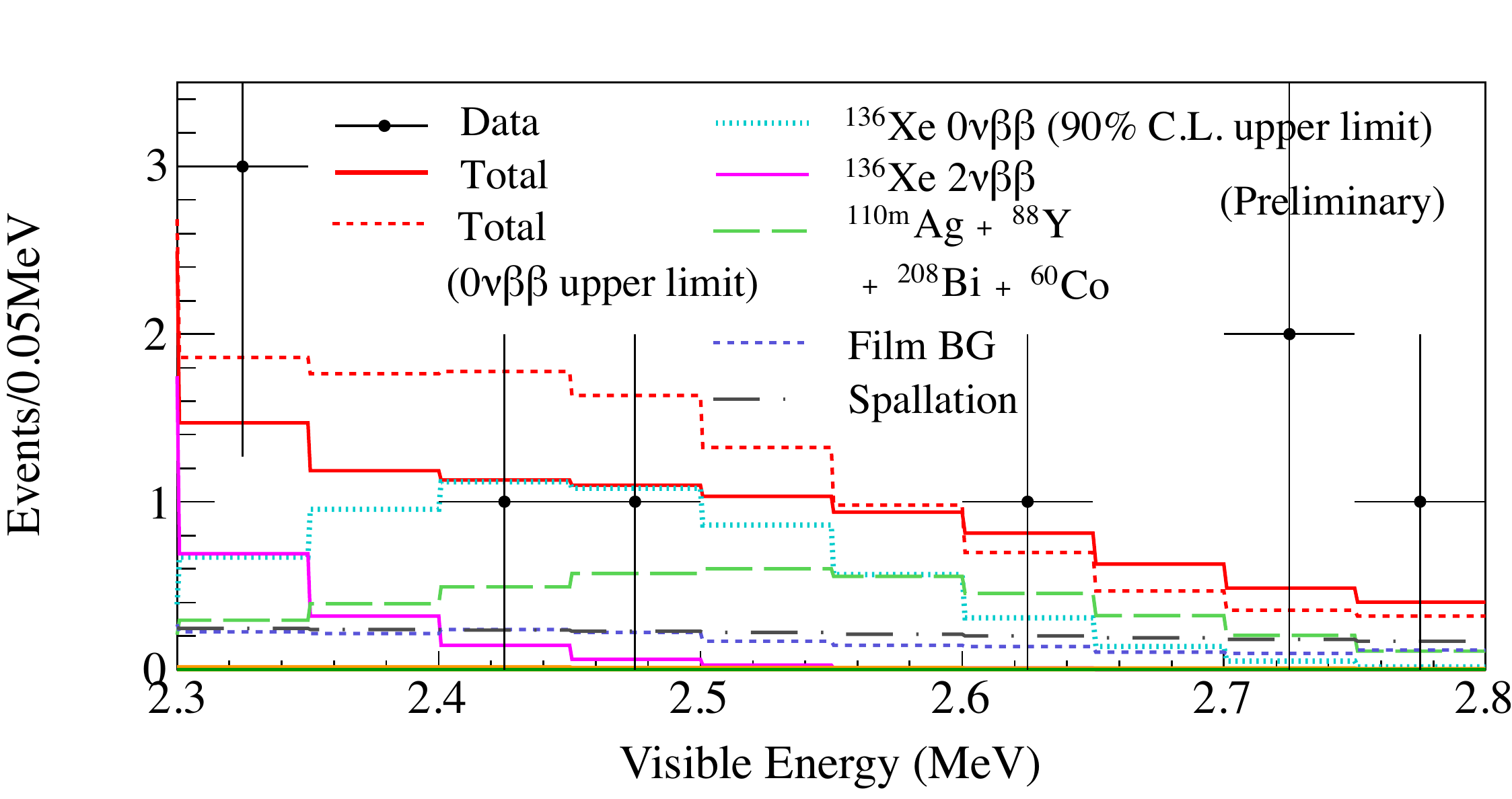}
  \caption{KamLAND-Zen Phase-II preliminary energy spectra of selected double-$\beta$ candidates within a radius
  cut of $R<1.0$ m \cite{ASA14}. }
%  \caption{Linear and expanded version of Fig. 7 showing the preliminary energy spectrum
%  of double-$\beta$ decay candidates within the radius cut of $R<1.0$ m shown together with the
%  best-fit backgrounds and the 90\% C.L. upper limit for $0\nu\beta\beta$ events \cite{ASA14}.}
  \label{fig8}
\end{figure}

A preliminary energy spectrum is shown in Fig.~\ref{fig8} for the internal volume ($R<1.0$ m) along
with the best fit background candidates. The potential background contributions of
$^{110m}$Ag,
$^{88}$Y, $^{208}$Bi, and $^{60}$Co in the 0$\nu\beta\beta$ visible energy region of interest are allowed
to vary in the fit. No events in excess over the background expectation were observed. The
90\% C.L. upper limit for a $0\nu\beta\beta$ contribution results in a life-time
limit of $T_{1/2}^{0\nu} > 1.3 \times 10^{25}$ yr. 

\begin{figure}[htb]
  \centering
  \includegraphics[width=2.5in]{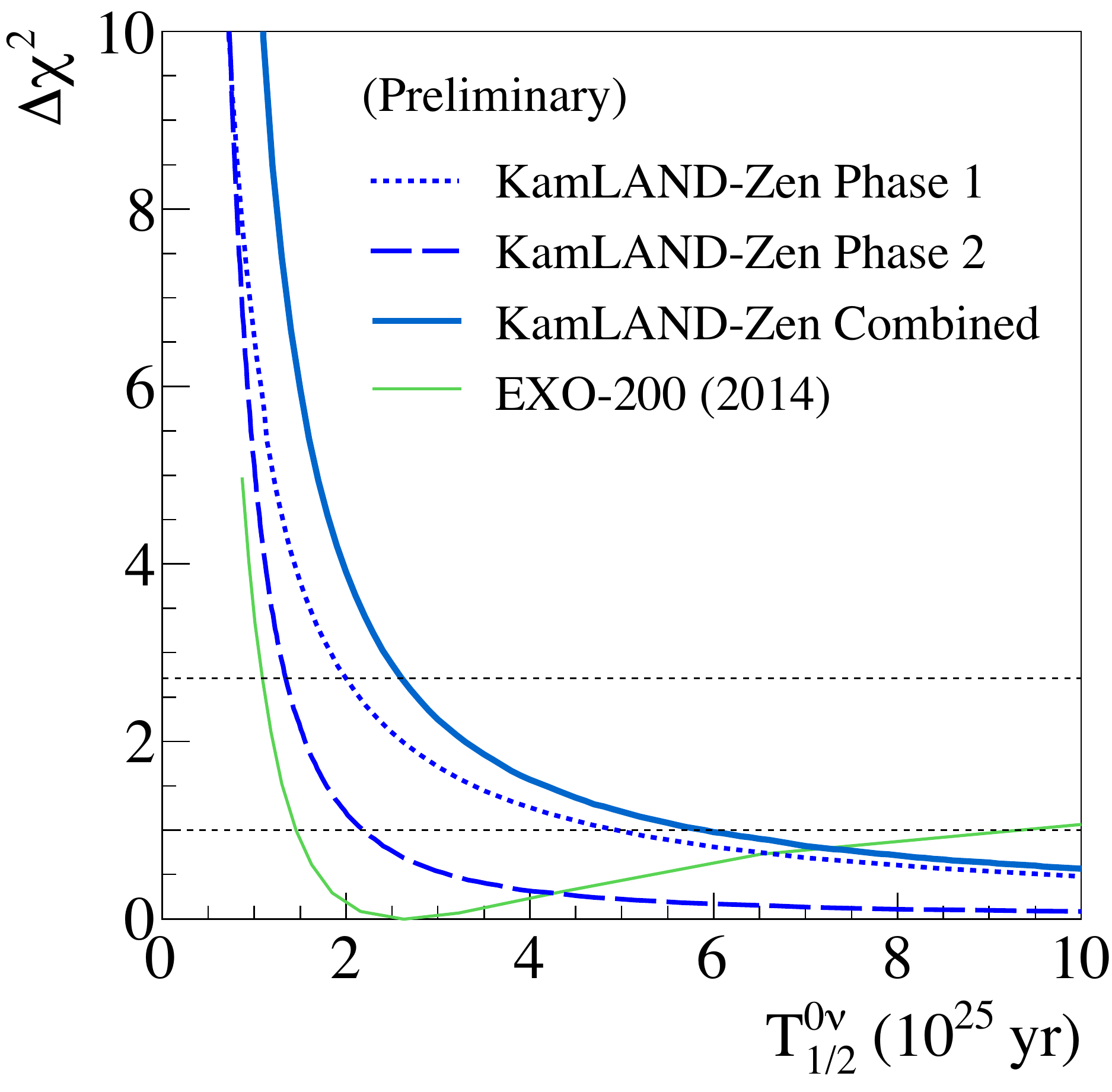}
  \caption{$\Delta\chi^2$ profile from fit to the half-life time of $^{136}$Xe
  $0\nu\beta\beta$ decay obtained in
  KamLAND-Zen Phase I \cite{GAN13}, KamLAND-Zen Phase II \cite{ASA14}, and the latter two results
  combined in comparison to EXO-200 \cite{ALB14}.}
  \label{fig9}
\end{figure}

Combining the KamLAND-Zen Phase I and Phase II results gives a 90\% C.L. lower limit of
$T_{1/2}^{0\nu} >2.6 \times 10^{25}$ yr. The individual KamLAND-Zen lower limits and the
combined KamLAND-Zen limit are displayed in Fig.~\ref{fig9} together with the EXO-200 result of
$T_{1/2}^{0\nu} > 1.1 \times 10^{25}$ yr \cite{ASA14}. The combined KamLAND-Zen half-life time
limit corresponds to an upper limit for the effective Majorana neutrino mass of
$\langle m_{\beta\beta}\rangle <(140-280)$ meV, using NME calculations of \cite{FAE12}. It is expected that
the $T_{1/2}^{0\nu}$ sensitivity will reach $3 \times 10^{25}$ yr (90\% C.L.) within two
years of data-taking (using Phase II data only).  

Various improvements are planned to increase the KamLAND-Zen sensitivity in the
foreseeable future \cite{ASA14}. First, a new and larger mini-balloon would increase the xenon
amount from presently 383 kg to 600 kg (or even 700 - 800 kg, if possible), and would
reduce the mini-balloon radioactivity by using ultra radio-pure materials. The associated
increase in fiducial Xe-LS mass will result in a sensitivity approaching $2 \times
10^{26}$ yr after
two years of data-taking. This sensitivity corresponds to $\langle m_{\beta\beta} \rangle = 50$ meV
using the largest of the NMEs of Ref. \cite{FAE12}. Second, to cover a large portion of the
inverted hierarchy or $\langle m_{\beta\beta}\rangle \approx 20$ meV sensitivity region requires a
substantial detector upgrade, referred to as “KamLAND2-Zen”. The associated proposal calls
for an improvement in energy resolution from currently 4\% to $<2.5$\% at $\approx2.5$ MeV
to better discriminate against the $2\nu\beta\beta$ ``background.'' This goal could be
accomplished by a combination of major modifications, like using a brighter scintillator,
installing higher quantum efficiency PMTs, and introducing light-collecting mirrors.
Furthermore, the amount of enriched xenon must be increased to 1,000 kg or more in order
to reach the goal of $\langle m_{\beta\beta}\rangle =20$ meV after five years of data-taking.

\end{document}